\documentstyle[preprint,prc,eqsecnum,aps,epsf]{revtex}

\begin{document}

\draft
\tightenlines

\title {Modern NN Force Predictions for the Total ND Cross Section up
to 300 MeV}

\author{H.~Wita{\l}a\footnote{Permanent address:
Institute of Physics, Jagellonian University, PL-30059 Cracow,
Poland}, H. Kamada\footnote{Present address:
Institut f\"ur Strahlen und Kernphysik der Universit\"at Bonn, Nussallee 14-16, D-53115 Bonn, Germany}, A. Nogga , W. Gl\"ockle}
\address{ Institute for Theoretical Physics II, Ruhr-University Bochum,
D-44780 Bochum, Germany.}

\author{ Ch.~Elster}
\address{Institute of Nuclear and
Particle Physics, and Department of Physics, \\ Ohio University,
Athens, OH 45701}

\author{D. H\"uber}
\address{
 Los Alamos National Laboratory, M.S.B283,  Los Alamos, NM 87544}


\date{\today}

\maketitle

\begin{abstract}
For several modern nucleon-nucleon potentials
state-of-the-art Faddeev calculations 
are carried out for the $nd$ total cross section between 10 and 
300~MeV projectile energy and compared to new high precision
measurements. The agreement between theory and data is rather good,
with exception at the higher energies where a 10\% discrepancy builds
up. In addition the convergence of the multiple scattering series
incorporated in the Faddeev scheme is studied numerically with
the result, that rescattering corrections remain important. 
Based on this multiple scattering series the high energy limit of the
total $nd$ cross section is also investigated  analytically. 
In contrast to the naive
expectation that the total $nd$ cross section is the sum of the 
$np$ and $nn$ total cross sections we find additional effects 
resulting from
the rescattering processes, which have different signs and different
behavior as function of the energy.  A shadowing effect in
the high energy limit only occurs for energies higher than 300~MeV.
The expressions in the high energy limit have qualitatively a similar
behavior as the exactly calculated expressions, but can be expected to 
be valid quantitatively only at much higher energies.

\end{abstract}

\vspace{5mm}


\pagebreak


\narrowtext


\section{Introduction}

 For three-nucleon scattering the total neutron-deuteron ($nd$) 
cross section is the 
 simplest observable, since it is integrated over the angular
 distribution in elastic $nd$ scattering and all the angular and
 continuous energy distributions of the 
three-nucleon (3N) breakup process. If 
 the theory does not agree with experiment, one has to expect that for 
 some individual observables even stronger discrepancies might exist.
 On the other hand if there is agreement, possible discrepancies in
 some individual unpolarized differential cross sections  have at
 least to average out. It is the
 aim of this article to compare precise new data between 10 and 300
 MeV neutron laboratory energy for the total $nd$ cross section
\cite{ref1} 
 with fully converged Faddeev calculations based on 
the most modern nucleon-nucleon (NN) forces. 
The calculations are based on a strictly nonrelativistic treatment. Despite 
 of this apparent restriction,
 we think that  this comparison of theory and experiment will be
 an important benchmark result. 

 A first presentation of our results appeared in \cite{ref1}. 
Here we want to give 
a more detailed description and explicitly show that our theoretical results 
are stable under the exchange of one of the most modern NN 
potentials by another one. 
Detailed studies on the size of the contributions of
 the first few terms in the  multiple scattering expansion to
 the total $nd$ cross section are given. 

In order to obtain more insight into the behavior of the multiple
scattering series for the total $nd$ cross section we provide analytical
high energy expansions for the first few terms. Compared to the most
naive picture, which in the high energy limit would equate the total
$nd$ cross section with the sum of the total cross sections for
neutron-proton ($np$) and neutron-neutron ($nn$) scattering, the rigorously
calculated result is smaller for energies up to 300~MeV. Obviously
shadowing effects can be expected. However, rescattering processes may 
{\it a priori} also enhance the total $nd$ cross section. It is shown
that rescattering processes of second order in the NN t-matrix lead to
shadowing and antishadowing effects. In principle these features are
known for the 3N scattering amplitude from studies in the framework of
Glauber theory \cite{glauber,joachain,henley,abers}. In contrast to this
formulation we start from a multiple scattering series, which implies
that it is an ordering according to powers of the NN t-matrix. The
analytical steps leading to the high energy limit of the first leading
terms are carried out in well defined integrals. There are no
{\it a priori} assumptions involved about the scattering process, such
as, e.g. diffraction type approximations. We observe that in the high
energy limit a shadowing effect only occurs for energies larger than 
300~MeV, below this energy the   contributions 
second order in the NN t-matrix in the high
energy limit enhance the sum of $np$ and $nn$ total cross sections.
Furthermore, it is interesting to see that even at 300~MeV the high
energy limit is quantitatively not yet reached. Nevertheless, the
analytical studies make the underlying physics more transparent. An
analogous investigation for the much simpler case of 3 bosons was
performed in Ref.~\cite{ref3}.

The paper is organized as follows. In Sec. II we briefly describe the
Faddeev framework, its multiple scattering expansion, the leading order
terms in the NN t-matrix for obtaining the $nd$ total cross section, and
the high energy limit of the corresponding expressions. The  derivation
of the high energy limit is explicitly given in Appendices A and B. In
Sec. III we present the results of our exact Faddeev calculations in
comparison with the data and with the calculations in the high energy
limit. In Sec. IV we study the effect of three-nucleon forces (3NF) on
the total $nd$ cross section. In addition, a simple estimate of
relativistic kinematic effects is given. We conclude in Sec. V.

\section{Leading Multiple Scattering Terms for the $nd$ Total Cross
Section and High Energy Limit}

We solve the Faddeev equations for 3N scattering in the form
\begin{equation}
 T| \Phi\rangle  = t P|\Phi \rangle +  t P G_0 T|\Phi\rangle,
\label{eq:1}
\end{equation}
where $t$ is the NN t-matrix, the operator $P$  the sum of a cyclical and an
anticyclical permutation of 3 objects and $ |\Phi \rangle $  
the initial channel
state composed of a deuteron and the momentum eigenstate of the
projectile nucleon. The full breakup operator is
\begin{equation}
 U_0  = (1 + P)T,
\label{eq:2}
\end{equation}
 defining the physical meaning of the $T$-operator. The
 iteration of Eq.~(\ref{eq:1})  together with 
 Eq.~(\ref{eq:2}) generates the multiple scattering
 series for the breakup process.

 Furthermore, the operator for elastic $nd$ scattering is
given by
\begin{equation}
 U  =  P G_0^{-1} +  P T .
\label{eq:3}
\end{equation}
 Our aim is the evaluation and the understanding of the total $nd$
 cross section, which is given via the optical theorem
\begin{equation}
 \sigma_{tot}^{nd}  = -(2\pi)^3 {{4m}\over{3q_0}}  
\: {\rm Im} \langle \Phi | U | \Phi \rangle .
\label{eq:4}
\end{equation}
 Here $m$ is the nucleon mass and $q_0$ is the asymptotic momentum of
the projectile nucleon relative to the deuteron.
Thus solving Eq.~(\ref{eq:1})  and  
using Eq.~(\ref{eq:3})  yields the desired
 result in the form Eq.~(\ref{eq:4}).

 In order to achieve insight how the total $nd$ cross section is formed 
 at high energies we consider the multiple scattering expansion 
of the operator $U$ as given in  Eq.~(\ref{eq:3}).
The first few terms constitute a power series
 expansion in the NN t-matrix
\begin{equation}
 U  =  P G_0^{-1} +  PtP + PtG_0PtP + \ldots .
\label{eq:5}
\end{equation}
At higher energies only a few terms in this expansion may be sufficient
to describe the total $nd$ cross section. We evaluate the different
terms in the series numerically and study their contributions to
the total cross section   in Sec.~III.

In order to obtain analytical insight into the contributions to
the total cross section we consider the imaginary part of the  operator 
$U$ as given in Eq.~(\ref{eq:5}) between the channel states 
$| \Phi \rangle$ and obtain 
\begin{eqnarray}
\label{eq:6}
\nonumber
 2i \: {\rm Im} \langle \Phi | U | \Phi \rangle &=& 
 \langle \Phi | U | \Phi \rangle - {\langle \Phi | U | \Phi \rangle}^* \\ 
&=&  \langle \Phi | P(t-t^{\dagger})P | \Phi \rangle + 
 \langle \Phi | PtG_0PtP - Pt^{\dagger}G_0^*Pt^{\dagger}P | \Phi \rangle 
+ \ldots
\end{eqnarray}
Since the first term in Eq.~(\ref{eq:5}) is real, it does not contribute
to the total cross section. 
 The second order term  in the NN t-matrix can be rewritten as
\begin{eqnarray}
\label{eq:7}
\nonumber
 \langle \Phi | PtG_0PtP - Pt^{\dagger}G_0^*Pt^{\dagger}P | \Phi \rangle &=& 
 \langle \Phi | P(t-t^{\dagger})PG_0tP | \Phi \rangle \\ 
 &-&  {\langle \Phi | P(t-t^{\dagger})PG_0tP | \Phi \rangle}^* 
 + \langle \Phi | Pt^{\dagger}P(G_0 - G_0^*)tP | \Phi \rangle,
\end{eqnarray}
 which will be the starting point for extracting the 
 high energy limit analytically.

 Let us first concentrate on the first order term of Eq.~(\ref{eq:6}) 
and work out the
 permutations. For the channel state $| \Phi \rangle$ we choose 
\begin{equation}
  | \Phi \rangle   = {| \Phi \rangle}_1 \equiv  
 {| \varphi_d \rangle}_{23} {| \bf{q_0} \rangle}_1
\label{eq:8}
\end{equation}
 and the NN t-matrix as given in the subsystem (23)
\begin{equation}
  t  \equiv t_1 \equiv t_{23} .
\label{eq:9}
\end{equation}

 The index 1 is the convenient and usual notation  \cite{ref4} to single out
 the pair (23). Similarily we shall denote $t_2 \equiv t_{31}$ and 
 $ t_3 \equiv t_{12}$.  
 The permutation operator $P$  is given as
\begin{equation}
  P  = P_{12}P_{23} +  P_{13}P_{23} .
\label{eq:10}
\end{equation}
Next we define
\begin{eqnarray}
\label{eq:11}
\nonumber
  {| \Phi \rangle}_2 &\equiv&   P_{12}P_{23}{ | \Phi \rangle}_1 \\ 
  {| \Phi \rangle}_3 &\equiv&   P_{13}P_{23}{ | \Phi \rangle}_1 
\end{eqnarray}

\noindent
 Applying P to the left and right one  obtains for
the first term on the right hand side of  Eq.~(\ref{eq:6}) 
\begin{eqnarray}
\label{eq:12}
\nonumber
 \langle \Phi | P(t-t^{\dagger})P | \Phi \rangle  &=& 
  ~_3{\langle} \Phi | t-t^{\dagger} | \Phi {\rangle}_2 + 
 ~_2{\langle} \Phi | t-t^{\dagger} | \Phi {\rangle}_2    \\ 
  &+&  ~_3{\langle} \Phi | t-t^{\dagger} | \Phi {\rangle}_3 + 
 ~_2{\langle} \Phi | t-t^{\dagger} | \Phi {\rangle}_3 .
\end{eqnarray}

 An immediate consequence of  Eq.~(\ref{eq:11}) and the 
antisymmetry of the deuteron
 state $| \varphi_d \rangle$ is
\begin{equation}
 {| \Phi \rangle}_3   = -P_{23}{| \Phi \rangle}_2 .
\label{eq:13}
\end{equation}
Consequently one has
\begin{eqnarray}
\label{eq:14}
\nonumber
  ~_3{\langle} \Phi | t-t^{\dagger} | \Phi {\rangle}_3  &=& 
  ~_2{\langle} \Phi | P_{23}(t-t^{\dagger})P_{23} | \Phi {\rangle}_2 = 
 ~_2{\langle} \Phi | t-t^{\dagger} | \Phi {\rangle}_2    \\ 
  ~_3{\langle} \Phi | t-t^{\dagger} | \Phi {\rangle}_2  &=& 
 - ~_2{\langle} \Phi | P_{23}(t-t^{\dagger}) | \Phi {\rangle}_2 = 
 ~_2{\langle} \Phi | t-t^{\dagger} | \Phi {\rangle}_3 .
\end{eqnarray}
Here
 we used the symmetry of $t$ under exchange of particles 2 and 3.

As final result we obtain for the first term on the right hand side of
Eq.~(\ref{eq:6}), the first  order term of the multiple scattering
expansion for the $nd$ total cross section
\begin{eqnarray}
\label{eq:15}
\nonumber
 {\langle}\Phi | P(t-t^{\dagger})P | \Phi {\rangle}  &=& 
 2 \: \lbrace  _2{\langle} \Phi | t-t^{\dagger} | \Phi {\rangle}_2 +  
 ~_2{\langle} \Phi | t-t^{\dagger} | \Phi {\rangle}_3 \rbrace   \\ 
  &=&  2 ~_2{\langle} \Phi | (t-t^{\dagger})(1-P_{23}) | \Phi {\rangle}_2 
\end{eqnarray}

 It is a straightforward but tedious algebra
 to evaluate  Eq.~(\ref{eq:15}) in the high
energy limit. The derivation is sketched in Appendix A.
Thus, the first order term in the multiple scattering expansion gives
 in the high energy limit 
the following contribution to the total $nd$ cross section
\begin{equation}
 {\sigma}_{nd,{\rm tot}}^{(1)} 
 = {\sigma}_{np}^{{\rm tot}} + {\sigma}_{nn}^{{\rm tot}}.
\label{eq:16}
\end{equation}

 The high energy limit is defined as the projectile momentum $q_0$
 being much larger than the typical momenta inside the deuteron.
Thus Eq.~(\ref{eq:16}) shows that in the high energy limit
the total cross section for $nd$ scattering in first order in the
NN t-matrix is indeed
given as the sum of the total cross sections for $np$ and $nn$
scattering.
This corresponds to the naive expectation, where the projectile
 nucleon scatters independently from both constituents of the deuteron.

Next we study the rescattering processes of second order in $t$  as
given in  Eq.~(\ref{eq:7}).
 Working out the permutations the first  term of the right hand side
of  Eq.~(\ref{eq:7}) can be written as 
\begin{eqnarray}
\label{eq:17}
\nonumber
 {\langle}\Phi | P(t-t^{\dagger})PG_0tP | \Phi {\rangle}  &=& 
 2 \: \lbrace  _2{\langle} \Phi | (t-t^{\dagger})G_0t_3( | \Phi {\rangle}_1 + 
 | \Phi {\rangle}_2 ) \\ 
 &+& 2 \: \lbrace  _2{\langle} \Phi | (t-t^{\dagger})G_0t_2 
(|\Phi {\rangle}_1 + | \Phi {\rangle}_3 ).
\end{eqnarray}
 Similarly the remaining term in  Eq.~(\ref{eq:7}) yields
\begin{eqnarray}
\label{eq:18}
\nonumber
 {\langle}\Phi | Pt^{\dagger}P(G_0-G_0^*)tP | \Phi {\rangle}  &=& 
 -4{\pi}i \: \lbrace  
 _2{\langle} \Phi | t^{\dagger}{\delta}(E-H_0)t_3( | \Phi {\rangle}_1 + 
 | \Phi {\rangle}_2 )  \\ 
 ~& & + \: _2{\langle} \Phi | t^{\dagger}{\delta}(E-H_0)t_2
 (|\Phi {\rangle}_1 + | \Phi {\rangle}_3 ) \rbrace .
\end{eqnarray}

The analytic evaluation of the high energy limit is even more tedious
and is 
sketched in Appendix B. The leading contribution of the terms given
in Eq.~(\ref{eq:7}), second
order in $t$ in the multiple scattering expansion 
 to the total $nd$ cross  section,  is given by

\begin{eqnarray}
\label{eq:19}
\nonumber
 {\sigma}_{nd,{\rm tot}}^{(2)} &=& -{1\over {4\pi}}{\sum_{ll'}}i^{l'-l}
      {\int_0^{\infty}}dr {\varphi}_l(r) {\varphi}_{l'}(r) \times \\
 \nonumber 
 & & ~~\lbrack 2~C(11l,1,-1)~C(11l',1,-1) \\ 
\nonumber
&~& \lbrace  
 -{1\over 2}({{4m}\over {3q_0}})^2(2\pi)^6 
( {1\over 2} ( {\rm Im} \langle -{3\over 4}{\bf q_0} {1\over 2}
 {1\over 2}np | t  | {3\over 4}{\bf q_0} {1\over 2}{1\over 2}np
{\rangle}_a )^2 \\ 
\nonumber
 & & - {1\over 2} ({\rm Re} \langle -{3\over 4}{\bf q_0} {1\over 2}
 {1\over 2}np | t  | {3\over 4}{\bf q}_0 {1\over 2}{1\over 2}np
{\rangle}_a )^2 \\
\nonumber
 & & +{1\over 2} ({\rm Im} \langle -{3\over 4}{\bf q_0} -{1\over 2}
 {1\over 2}np | t  | {3\over 4}{\bf q}_0 {1\over 2}-{1\over 2}np
{\rangle}_a )^2 \\
\nonumber
 & & - {1\over 2} ({\rm Re} \langle -{3\over 4}{\bf q_0} -{1\over 2}
 {1\over 2}np | t  | {3\over 4}{\bf q}_0 {1\over 2}-{1\over 2}np
{\rangle}_a )^2 \\
\nonumber
 & & + {\rm Re} \langle {3\over 4}{\bf q_0} {1\over 2}
 {1\over 2}np | t  | {3\over 4}{\bf q}_0 {1\over 2}{1\over 2}np
{\rangle}_a 
 {\rm Re} \langle {3\over 4}{\bf q_0} {1\over 2}
 {1\over 2}nn | t  | {3\over 4}{\bf q}_0 {1\over 2}{1\over 2}nn
{\rangle}_a  \\
\nonumber
 & & + {\rm Re} \langle {3\over 4}{\bf q_0} {1\over 2}
 -{1\over 2}nn | t  | {3\over 4}{\bf q}_0 {1\over 2}-{1\over 2}nn
{\rangle}_a 
 {\rm Re} \langle {3\over 4}{\bf q_0} {1\over 2}
 -{1\over 2}np | t  | {3\over 4}{\bf q}_0 {1\over 2}-{1\over 2}np
{\rangle}_a ) \rbrace \\
\nonumber 
 & & + C(11l,00)~C(11l',00) \\ 
\nonumber
&~& \lbrace  - {1\over 2}({{4m}\over {3q_0}})^2(2\pi)^6 
( {1\over 2} ({\rm Im} \langle -{3\over 4}{\bf q_0} {1\over 2}
 -{1\over 2}np | t  | {3\over 4}{\bf q_0} {1\over 2}-{1\over 2}np
{\rangle}_a )^2 \\ 
\nonumber
 & & - {1\over 2} ({\rm Re} \langle -{3\over 4}{\bf q_0} {1\over 2}
 -{1\over 2}np | t  | {3\over 4}{\bf q}_0 {1\over 2}-{1\over 2}np
{\rangle}_a )^2 \\
\nonumber
 & & + {\rm Im} \langle -{3\over 4}{\bf q_0} {1\over 2}
 {1\over 2}np | t  | {3\over 4}{\bf q}_0 {1\over 2}{1\over 2}np
{\rangle}_a 
 {\rm Im} \langle -{3\over 4}{\bf q_0} -{1\over 2}
 {1\over 2}np | t  | {3\over 4}{\bf q}_0 {1\over 2}-{1\over 2}np
{\rangle}_a  \\
\nonumber
 & & - {\rm Im} \langle {3\over 4}{\bf q_0} {1\over 2}
 -{1\over 2}np | t  | {3\over 4}{\bf q}_0 -{1\over 2}{1\over 2}np
{\rangle}_a 
 {\rm Im} \langle {3\over 4}{\bf q_0} {1\over 2}
 -{1\over 2}nn | t  | {3\over 4}{\bf q}_0 -{1\over 2}{1\over 2}nn
{\rangle}_a  \\
\nonumber
 & & + {\rm Re} \langle {3\over 4}{\bf q_0} {1\over 2}
 {1\over 2}np | t  | {3\over 4}{\bf q}_0 {1\over 2}{1\over 2}np
{\rangle}_a 
 {\rm Re} \langle {3\over 4}{\bf q_0} {1\over 2}
 -{1\over 2}nn | t  | {3\over 4}{\bf q}_0 {1\over 2}-{1\over 2}nn
{\rangle}_a  \\
\nonumber
 & & + {\rm Re} \langle {3\over 4}{\bf q_0} {1\over 2}
 {1\over 2}nn | t  | {3\over 4}{\bf q}_0 {1\over 2}{1\over 2}nn
{\rangle}_a 
 {\rm Re} \langle {3\over 4}{\bf q_0} {1\over 2}
 -{1\over 2}np | t  | {3\over 4}{\bf q}_0 {1\over 2}-{1\over 2}np
{\rangle}_a  \\
\nonumber
 & & - {\rm Re} \langle -{3\over 4}{\bf q_0} {1\over 2}
 {1\over 2}np | t  | {3\over 4}{\bf q}_0 {1\over 2}{1\over 2}np
{\rangle}_a 
 {\rm Re} \langle -{3\over 4}{\bf q_0} -{1\over 2}
 {1\over 2}np | t  | {3\over 4}{\bf q}_0 {1\over 2}-{1\over 2}np
{\rangle}_a  \\
\nonumber
 & & +{\rm Re} \langle {3\over 4}{\bf q_0} {1\over 2}
 -{1\over 2}np | t  | {3\over 4}{\bf q}_0 -{1\over 2}{1\over 2}np
{\rangle}_a 
 {\rm Re} \langle {3\over 4}{\bf q_0} {1\over 2}
 -{1\over 2}nn | t  | {3\over 4}{\bf q}_0 -{1\over 2}{1\over 2}nn
{\rangle}_a ) \rbrace \rbrack \\
\nonumber
&-& {1\over {4\pi}}{\sum_{ll'}}i^{l'-l}
{\int_0^{\infty}}dr {\varphi}_l(r) {\varphi}_{l'}(r) \times \\
\nonumber
 & & ~~ \lbrack 2~C(11l,1,-1)~C(11l',1,-1) \\
\nonumber
&~& \lbrace {1\over 2} ({\sigma}_{np}^{{\rm tot}}({1\over 2}{1\over 2})
              {\sigma}_{nn}^{{\rm tot}}({1\over 2}{1\over 2}) +
              {\sigma}_{np}^{{\rm tot}}({1\over 2}-{1\over 2})
              {\sigma}_{nn}^{{\rm tot}}({1\over 2}-{1\over 2})  )  \\
\nonumber
 & & + C(11l,00)~C(11l',00) \\
\nonumber
&~& \lbrace  {1\over 2} (  {\sigma}_{np}^{{\rm tot}}({1\over 2}{1\over 2})
              {\sigma}_{nn}^{{\rm tot}}({1\over 2}-{1\over 2}) +
              {\sigma}_{np}^{{\rm tot}}({1\over 2}-{1\over 2})
              {\sigma}_{nn}^{{\rm tot}}({1\over 2}{1\over 2})  )  
 \rbrace \rbrack \\
&\equiv&  \sigma^{(2)}_{I} ~ + ~ \sigma^{(2)}_{II} \\
\nonumber
\end{eqnarray}
\noindent
Here the sum over l, l' denote the S and D-wave contributions of the 
deuteron wave functions  ${\varphi}_l(r)$.  
Further occur the imaginary and real 
parts of the forward and backward two-nucleon  scattering amplitudes 
($nn$ or $np$) with specified spin magnetic quantum numbers in the initial 
and final states. The index a denotes  antisymmetrization without 
the factor ${1 \over \sqrt{2}}$. 
The terms of second order in $t$ give rise to Eq.~(\ref{eq:19}), 
a quite complicated expression 
exhibiting the interferences due to spin and isospin degrees of freedom. 
The part $\sigma^{(2)}_{I}$ related to the first square bracket in
Eq.~(\ref{eq:19}) contains  
real and imaginary
parts of the NN scattering amplitudes and has positive as well as
negative contributions. Only a numerical evaluation of these terms
as function of the scattering energy can give insight about the size and
energy dependence of these terms. The part $\sigma^{(2)}_{II}$ related to the second square bracket contains due  the optical theorem
a product of two NN total cross sections. That part $\sigma^{(2)}_{II}$ is negative  and  
reduces the term of first order in the multiple scattering
expansions, the sum of the two NN total cross sections. 
This is  naturally called shadowing effect.

Neglecting spin and isospin dependencies and consequently 
the D-wave admixture in the deuteron,  and further neglecting the backward 
amplitudes that expression given in Eq.~(\ref{eq:19}) reduces to the much simpler ones which are 
given in  \cite{ref3} for 3 bosons.

\section{Discussion of the Results of the Faddeev calculations and
of the High Energy Limit}

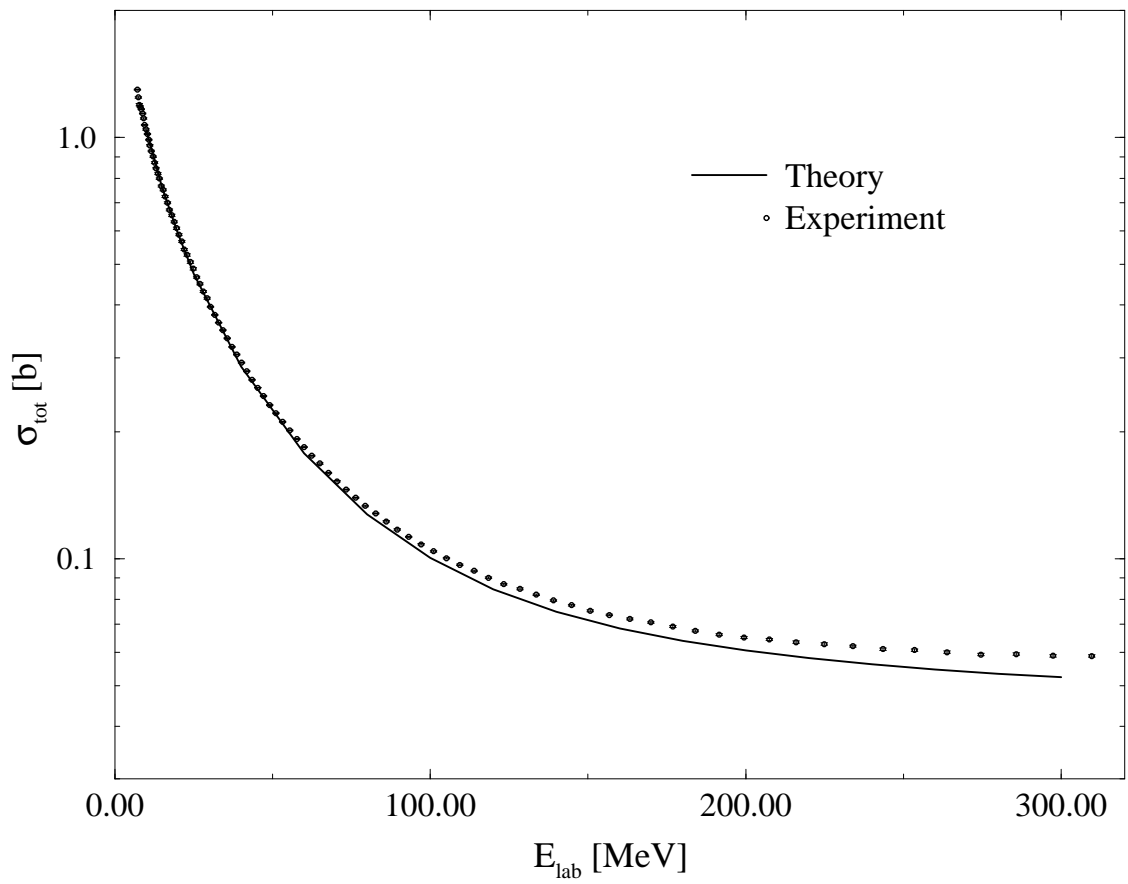
\begin{figure}
\caption{Comparison of the Faddeev calculation for the total
$nd$ cross section based on the CD-Bonn \protect\cite{ref6} potential with 
data \protect\cite{ref1}. The experimental values for the total
cross section for deuterium were obtained by adding the separately
measured values of the hydrogen cross section and the 
deuterium-hydrogen cross section difference (for details see
Ref.~\protect\cite{ref1}). Error bars are omitted,
since they are smaller than the dot size. \label{fig1}}
\end{figure}

For calculating the $nd$ total cross section we employ 
the most recent NN potentials, namely CD-Bonn \cite{ref6}, 
AV18 \cite{ref7} ,Nijm I and II  \cite{ref8}.
Those potentials are
optimally fitted to the Nijmegen data basis up to 350 MeV
nucleon laboratory energy. 
The Faddeev equation  Eq.~(\ref{eq:1}) is solved in
momentum space in a partial wave decomposed form. For a description of
the technical details of the calculations we refer to Ref.~\cite{ref5}.
In Fig.~\ref{fig1} we compare the calculated $nd$ total cross section based on
the CD-Bonn potential with the data. The experimental data are obtained
by adding the separately measured values of the hydrogen cross section
and the deuterium-hydrogen cross section difference \cite{ref1}. 
The figure shows that the theoretical calculation describes the data
very well at the lower energies but falls below the data at higher
energies. The calculation begins to underestimate the data around 
100~MeV by about 4\%, and this discrepancy increases to about 11\% at
300~MeV. 

The calculation presented in Fig.~\ref{fig1} employs in the 
two-nucleon subsystem
angular momenta up to $j_{max}$=4. This is sufficient if we are
satisfied with a calculational accuracy of 1\%, which corresponds to
the size of the experimental error. In order to demonstrate the
dependence of the numerical accuracy of our calculations, we show in
Table~\ref{tab1} for a few energies in the relevant energy regime the 
convergence of the total cross section as function of
$j_{max}$, the maximum angular momentum of the two-nucleon subsystem. 
 In the 3N calculations
two-nucleon angular momenta larger than $j_{max}$ are put to zero.
We see that $j_{max} =4$ is sufficient,
 if we are satisfied with an accuracy of 1\%. 
Next we
 demonstrate the stability of our theoretical result under the
 exchange of the NN potentials. This is shown in Table~\ref{tab2}. Clearly
within an accuracy of 1\% the predictions of those four essentially
 phase equivalent potentials agree with each other. From this we can
conclude that the deviation of the calculation from the data at higher
energies, which is shown in Fig.~\ref{fig1} is independent from the
NN interaction employed. 

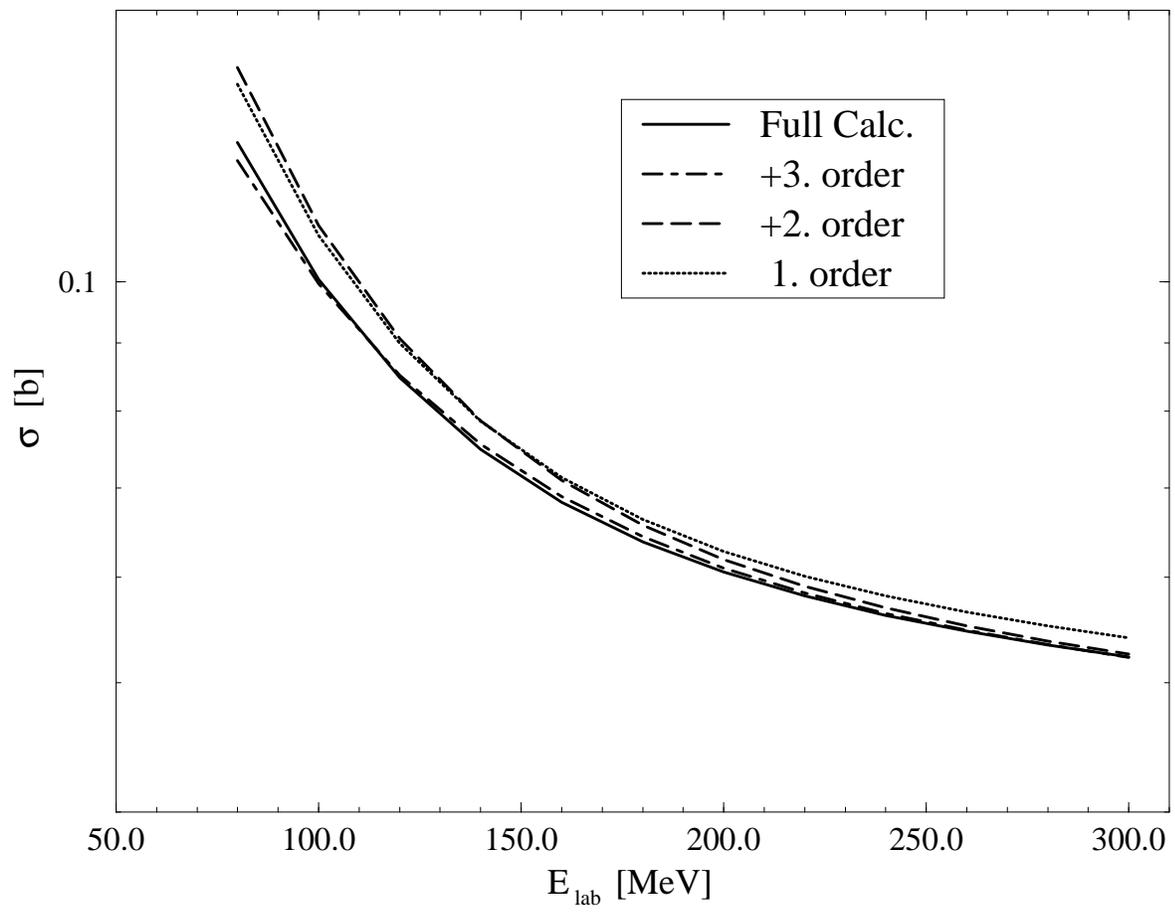
\begin{figure}
\caption{The contributions to the total $nd$ cross section 
of the different orders in the multiple
scattering expansion of the Faddeev amplitude. Successive orders are added
to the first order term and then compared with the full Faddeev
calculation. The potential employed is the CD-Bonn model.
\label{fig2}}
\end{figure}

In order to gain some insight how the total $nd$ cross section is build up,
the individual contributions of first, second
 and third order in the multiple scattering expansion of the elastic
 amplitude are compared. In Fig.~\ref{fig2} we present the total cross 
section calculated in first order, successively add the contributions
of the second and third order and compare those results with the full
calculation. The figure shows that the first order in the multiple
scattering expansion is larger than the full result at all energies.
The second order contribution enhances the cross section below about
130~MeV and decreases it at the higher energies. 
We also see that at energies below about
 200~MeV the third order term is significantly larger than the second
 order term, and  thus that rescattering of higher order is very important. 
Only above  200~MeV  the third order contribution
becomes negligible. Here the small correction of the second order
 rescattering process becomes sufficient to describe together with the
first order contribution the full solution of the Faddeev equation.
 Around 300~MeV the first two terms in the multiple scattering expansion
of the scattering amplitude are sufficient to describe the  total cross
section.

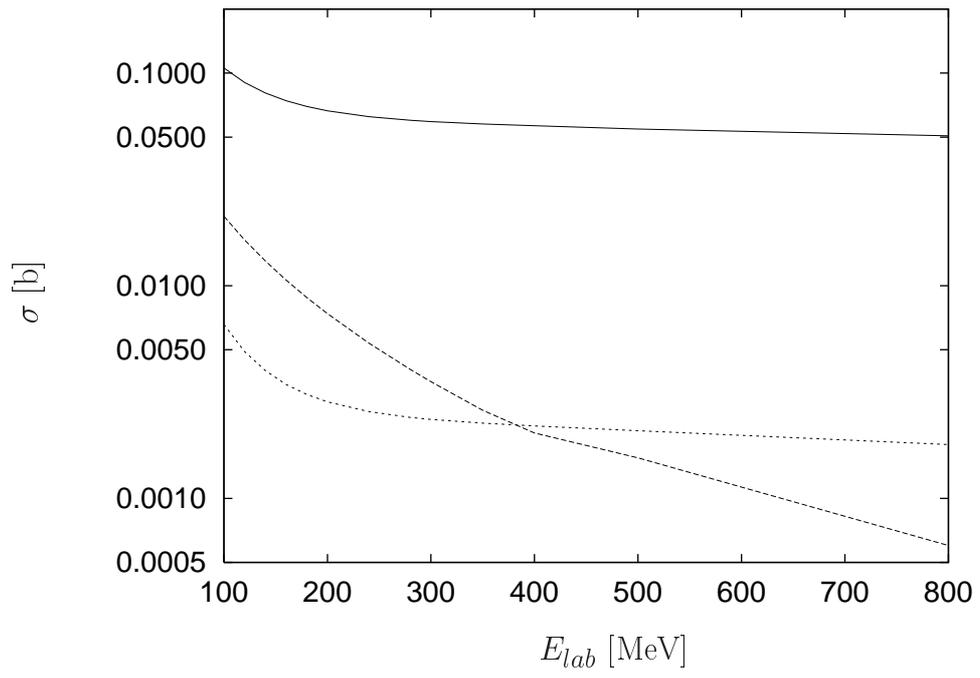
\begin{figure}
\caption{The contributions of the different orders in the NN t-matrix in
the high energy limit to the total $nd$ cross section. 
The solid  line shows the sum of the $np$ and $nn$ total cross
sections. 
The positive contribution $\sigma_{I}^{(2)}$ (dashed line) drops
faster in energy than   
the negative contribution $\sigma_{II}^{(2)}$ (its magnitude is shown
as dotted line)
of the second order term of the multiple scattering. For details see
Sec. III.  \label{fig3}}
\end{figure}

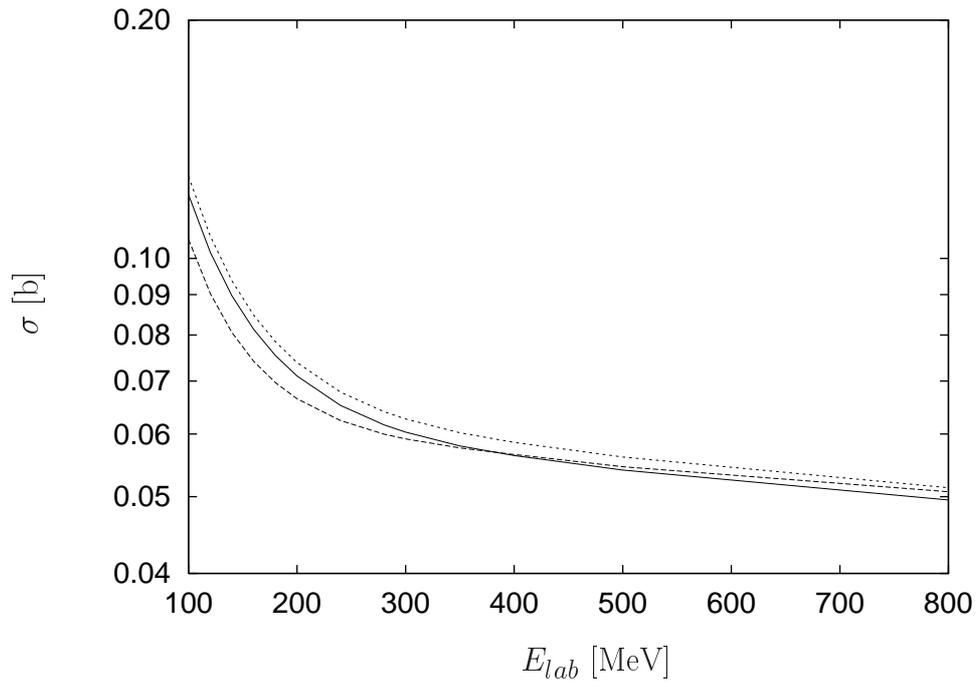
\begin{figure}
\caption{The contributions to the total $nd$ cross section  in
the high energy limit. 
The dashed   line shows the sum of the $np$ and $nn$ total cross
sections. Successively added to this is the positive contribution 
$\sigma_{I}^{(2)}$ (dotted line) and 
the negative contribution $\sigma_{II}^{(2)}$ (solid line).  
For details see
Sec.~III.  \label{fig4}}
\end{figure}

As a next step we consider the high energy limit of the first and
second order terms of the multiple scattering expansion as derived
in Sec. II. The first order term results in a sum of the total cross
section for $np$ and $nn$ scattering. The high energy limit of the
second order term is given in Eq.~(\ref{eq:19}). In this equation we
already indicate that this term consists of two parts, one which  
consists of products of real and imaginary
parts of the NN t-matrix, and one which is due to the optical theorem 
proportional to products of NN total cross sections. The latter term
is 
negative and can be
considered as shadowing effect. The contributions of the different terms
of the high energy limit are shown in Fig.~\ref{fig3}. 
We show the individual contributions up to a laboratory energy of
800~MeV in   order
to get a better insight into their energy dependence.  The
contribution of the sum of $\sigma_{np}+\sigma_{nn}$ is by far
dominant. The calculation shows that $\sigma_{I}^{(2)}$ is always
positive and thus enhances the total cross section. Thus this
term could be viewed as anti-shadowing effect. 
The term $\sigma_{II}^{(2)}$  is always negative.
We see that $\sigma_{I}^{(2)}$ falls off faster as function of
the energy than $\sigma_{II}^{(2)}$. This interplay of
the different contributions to the high energy limit of the second order
term in the multiple scattering expansion leads to an enhancement of
the total $nd$ cross section below about 400~MeV and a weakening above
compared to the sum of the $np$ and $nn$ total cross sections. Thus only
above 400~MeV the second order term in the elastic amplitude leads
 in the high energy limit to
a shadowing effect as it is familiar from older estimates within the
framework of Glauber theory. This is shown in Fig.~\ref{fig4}. 
The dashed line
gives the contribution of the sum of $\sigma_{np} + \sigma_{nn}$. To
this sum the term  $ \sigma_{I}^{(2)}$ is added and given by the
dotted line and enhances the total cross section. Then we add in
addition $\sigma_{II}^{(2)}$ leading to the solid line.    

It is now interesting to compare those results obtained from the first
two terms of the multiple scattering expansion in the high energy limit
with the corresponding contributions calculated exactly with the Faddeev
framework. This comparison is summarized in Table~\ref{tab3}. 
The second
column gives the exact Faddeev result as reference. It should be noted
that all results given in Table~\ref{tab3} are calculated using
$j_{max}$=4 in the two-nucleon subsystem. Starting at 100~MeV we compare
the exact first order result to its high energy limit, the sum of
the $np$ and $nn$ total cross sections,  as given in
Eq.~(\ref{eq:16}). The table shows that this sum is larger
than the exact first order term  for 140 MeV and higher. 
In fact, the difference increases with
increasing energy. In the case of three bosons interacting via
Malfliet-Tjon type forces (acting in {\bf all} partial waves) this
difference vanishes around 300~MeV and the asymptotic expression
acquires its validity here \cite{ref3}. Apparently, the three nucleon
system with realistic forces behaves differently. It would be very
interesting to find out at which energy the first order term will reach
the asymptotic form. This interest is however restricted to the context
of a mathematical model of potential scattering, since the standard NN
forces loose their physical meaning above pion-production threshold. 

Next we compare the exact second order result to the expression of the
high energy limit as given in Eq.~(\ref{eq:19}).  First, the exact
second order result of the Faddeev formalism is a small correction to
the first order result. At lower energies it is additive and changes
sign around 150~MeV, which then leads to a reduction of the cross
section.
As can be seen in Fig.~\ref{fig3}, the second order term in the high
energy limit exhibits this behavior only around 400~MeV. Comparing
the sum of first and second order terms of the exact Faddeev calculation
to the corresponding term in the high energy limit (column 9  which
is the sum of the terms in columns 6-8   in Table~\ref{tab3}) 
shows that the high energy limit is always larger
in the energy region under consideration. 
At 300~MeV the  difference  for the combined results in those orders
is still 8\%, and thus the high energy limit
is not yet quantitatively valid.

As remark we would like to add that the asymptotic form requires higher
NN force components than $j_{max}$=4. For the study presented in 
Table~\ref{tab3} we kept the angular momenta in the two-nucleon system
fixed at this value in order to compare with the Faddeev results. 
The two-nucleon $t$ matrices underlying 
the calculations presented in Fig.~\ref{fig3}, however, 
employ up to $j_{max}$=15
in order to obtain a converged result at 800~MeV. In the case of the
high energy limit this can be easily achieved, since only the NN
on-shell amplitudes enter.

\section{Estimate of Three-Nucleon Force and Relativistic Corrections}

In this section we want to study the effect of three nucleon forces
(3NF) on the total $nd$ cross section and give a very simple estimate on
the size of relativistic effects. The inclusion of  3NF in 3N continuum
Faddeev calculations has been formulated in Ref.~\cite{3NF1},
superseding a previous formulation given in Ref.~\cite{ref5}. The newer
formulation is formally more elegant and numerically more efficient and is 
used in our calculations. The inclusion of 3NF generalizes Eq.~(\ref{eq:1})
to
\begin{eqnarray}
\nonumber
 T| \Phi\rangle  &=& t P|\Phi \rangle + (1+t G_0) V_4^{(1)} (1+P)|\Phi
\rangle \\
& & + \ t P G_0 T|\Phi\rangle + (1+t G_0) V_4^{(1)} (1+P) G_0
T|\Phi\rangle, \label{eq:4.1}
\end{eqnarray}
 where $V_4^{(1)}$ is that part of the 3NF, which is symmetric under
exchange of particles $2$ and $3$. The total 3NF then has the form
\begin{equation}
V_4 = V_4^{(1)} + V_4^{(2)} + V_4^{(3)}, \label{eq:4.2} 
\end{equation}
and is totally symmetric. The full breakup operator as given in
Eq.~(\ref{eq:2}) keeps its form, whereas the operator for elastic $nd$ 
scattering changes to
\begin{equation}
 U  =  P G_0^{-1} + V_4^{(1)}(1+P) +  P T + V_4^{(1)}(1+P) G_0 T.
\label{eq:4.3}
\end{equation}
The optical theorem given in Eq.~(\ref{eq:4}) stays of course valid.

The numerical evaluation of Eq.~(\ref{eq:4.1}) is quite demanding, since
in contrast to the 3N bound state one needs the partial wave projected
momentum space representation of $V_4^{(1)}$ now for both parities and,
in addition, for total 3N angular momenta larger than $J=1/2$. At a
projectile energy $E_{lab}^n = 200$~MeV we checked that total angular
momenta up to $J=9/2$  were needed. The calculation of the
three-nucleon force matrix elements for such high angular momenta
became possible due to the new partial wave decomposition for the
three-nucleon force introduced in \cite{huber97}. 
Of course, NN forces contribute
significantly up to $J=25/2$ to 3N scattering states. In our
calculations we employ the Tucson-Melbourne force \cite{TM}, which has
has been adjusted individually to the $^3$H binding energy for each of
the different modern NN forces \cite{3NF2}.   The results presented
in this paper are based on the CD-Bonn potential and the corresponding
parameters of the Tucson-Melbourne force. 

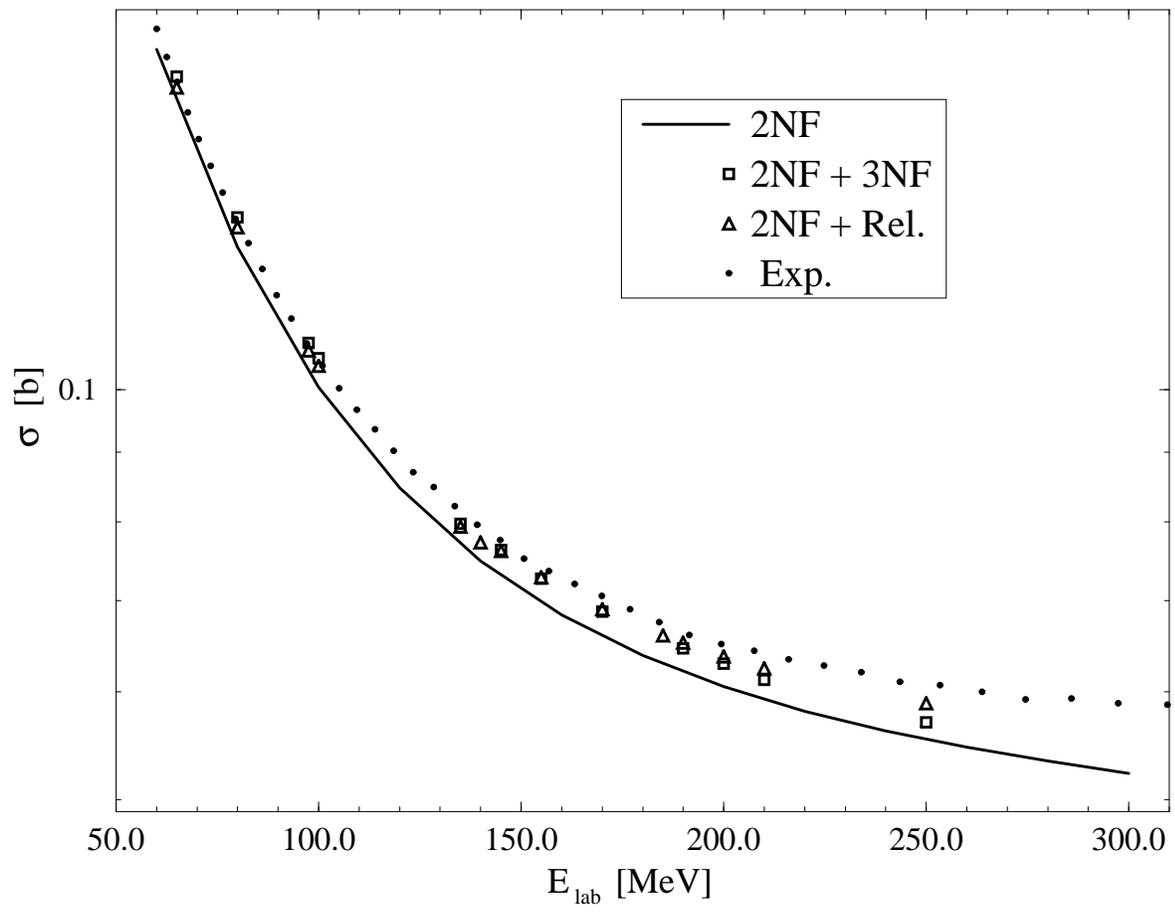
\begin{figure}
\caption{Corrections to the Faddeev calculation based on a nonrelativistic
Hamiltonian using strictly two-nucleon forces (solid line) for 
the total §nd§ cross section. The open squares show calculations at
various energies, where the Tucson-Melbourne 3N force has been included.
The open triangles show at various energies 
calculations based on two-nucleon forces corrected with a 
relativistic kinematic (for details see Sec. IV). 
The dots describe the data \protect\cite{ref1}.
All calculations are based on the CD-Bonn potential.  \label{fig5}}
\end{figure}

In a previous study we found that 3N force effects are responsible for
filling the minima in the angular distribution of elastic $nd$
scattering, especially at higher energies \cite{3NF3}. 
Though the effect is seen only in the minima of the differential cross
sections, and thus is small in magnitude, one might expect traces
thereof in the integrated quantity, the total cross section.
Corresponding effect to those in the elastic angular distribution might
also occur in 3N breakup processes, over which is also integrated when
calculating the total cross section. This is indeed the case as shown in
Fig.~\ref{fig5}. The effect of the 3N force enhances the total $nd$ 
cross section calculated with NN forces only by about 4\%. It is
interesting to note that this enhancement is almost independent of the
energy when considering the energy regime between 100 and 300~MeV. 
It is also obvious from the figure that the 3N force effects included
here are not strong enough to bridge the gap between the calculation
based on 2N forces only and the data. However, one should keep in mind
that the Tucson-Melbourne 3NF model is just one model, and theory as
well as application of 3N forces is just at the beginning. 

A second effect, which can be expected to appear especially at higher
energies is an effect due to relativity. A generally accepted framework
for carrying out relativistic 3N scattering calculations does not yet
exist. Thus we restrict ourselves to a very simple estimate of a
relativistic kinematic effect. Due to the optical theorem the total 
cross section is given as
\begin{equation}
 \sigma_{tot}^{nd}  \propto \frac{1}{|j|} {\rm Im} \langle \Phi | U | \Phi
\rangle, \label{eq:4.4}
\end{equation}
where $|j|$ is the incoming current density. If we neglect possible
changes of ${\rm Im} \langle \Phi | U | \Phi \rangle$ in going from a 
nonrelativistic to a relativistic formulation, the ratio of the
relativistic to the nonrelativistic total cross section is simply given
by
\begin{equation}
\frac{\sigma_{{\rm tot},rel}^{nd}}{\sigma_{{\rm tot},nr}^{nd}} =
 \frac{|j|_{nr}}{|j|_{rel}} 
 =  \frac{E_n E_d}{p_n^{rel} (E_n + E_d)} / 
          \frac{m_n m_d}{p_n^{nr} (m_n + m_d) }.
\label{eq:4.5}
\end{equation}
Here, $E_n$ and $E_d$ are the  relativistic kinetic energies 
of the neutron and the
deuteron in the c.m. system, and $p_n^{rel}$ is the relativistic 
c.m. momentum. The other terms are obviously the nonrelativistic
approximations thereof. This ratio given in Eq.~(\ref{eq:4.5})  is
easily evaluated and leads to an increase of the total $nd$ cross
section, namely an increase of about 3\% at 100~MeV and about 7\% at
250~MeV. Calculations based on 2N forces only, which include this 
relativistic effect, are shown at various  energies in Fig.~\ref{fig5}.
Again, these additional effects enhance the total cross section and
bring the calculation closer to the data. It is also clear that this
effect does not constitute a relativistic theory, since the forward
scattering amplitude is still calculated entirely nonrelativistically.
However, both effects, the one caused by inclusion of a 3NF model and 
the simple estimate of a relativistic effect, are roughly similar in
magnitude, and when added up would come close to the data in the
higher energy regime. It should be also noted that at lower energies,
the added effects would overpredict the data.

\section{Summary}

In view of new precise experimental information on the total $nd$ cross
section fully converged Faddeev calculations based on the most modern NN
forces are carried out at projectile energies between 10 and 300~MeV.
For the Faddeev calculations a strictly nonrelativistic treatment is
employed. The calculations show that the results for the total $nd$ cross
section do not depend on the choice among the most recent
phase-equivalent potentials. 

In  order to obtain more insight into the behavior of the multiple
scattering series for the $nd$ total cross section, we study its
convergence within the Faddeev framework. Below about 200~MeV projectile
energy rescattering of higher order is very important, as can be
concluded from the fact that here the third order contribution of the
multiple scattering series is significantly larger in magnitude than the
second order. Only around 300~MeV the first two terms in the multiple
scattering expansion are sufficient to describe the total $nd$ cross
section.

In order to obtain analytical insight we investigate the first terms of
the multiple scattering series for the forward elastic $nd$ scattering
amplitude resulting from the Faddeev equations in the high energy
limit. Although the treatment is purely nonrelativistic and enters far
into the region where relativity is expected to be important, we think
it is interesting to know the asymptotic behavior at high energies for
pure potential models without absorption. Absorption processes (particle
production) occurring in a relativistic context will change the results
presented here. In accordance with the naive expectation, we extract
from the first order term in the multiple scattering series the sum of
the total cross sections for $np$ and $nn$ scattering, and from the
second order term a shadowing effect proportional to products of the
total NN cross sections. The shadowing contribution is negative and
 reduces the total $nd$
cross section. However, we also find positive terms proportional to
products of real and imaginary parts of spin dependent NN forward and
backward scattering amplitudes. Those terms provide an overall
enhancement of the $nd$ total cross section, but vanish faster with energy
  than the
shadowing term. At projectile energies above 400~MeV the shadowing is
the dominant second order effect.

A comparison of the theoretical calculations based on a nonrelativistic
Faddeev framework with two-nucleon forces only and the experimental
observables exhibits a discrepancy with respect to the $nd$ data, which
starts around 100~MeV with a few percent and reaches about 10\% at
300~MeV. Possible corrections can be due to either three-nucleon forces
or relativistic effects or both . We calculated the effect of the
Tucson-Melbourne 3N force on the total $nd$ cross section and find that
in the energy regime between 100 and 300~MeV this 3NF model provides an
overall enhancement of the total cross section of $\sim$4\%. This is
consistent with previous studies, which found that the minima of the
 elastic scattering angular distribution are filled in when this 3NF is taken into
consideration. In the total cross section this effect is not large
enough to bring the present calculations close to the data, but the
trend is in  the right direction.

A second effect, which can be expected to appear especially at higher
energies is due to relativity. We carry out a simple estimate of
effects due to 
relativistic kinematics , and find that these corrections enhance
the total $nd$ cross section about 7\% at 250~MeV, while being much  smaller
at lower energies. Again, those corrections go into the right direction
with respect to the data. However, we do not suggest that our estimate
is the complete solution of the problem, since in our calculation the
forward scattering amplitude is still calculated entirely
nonrelativistically.

Both our estimates of corrections to our nonrelativistic Faddeev
calculations based on 2N forces  suggest, that effects due to 3N forces
as well as relativistic effects become non negligible at higher energies.
This calls for a strong effort to progress theoretical developments of
a theory of 3N forces as well as a relativistic framework for 3N
scattering.

\vfill

\acknowledgments

This work was performed in part under the auspices of the U.~S.
Department of Energy under contracts No. DE-FG02-93ER40756 with Ohio
University, the NATO Collaborative Research Grant 960892, the National
Science Foundation under Grant No. INT-9726624 and the Deutsche
Forschungsgemeinschaft. The work of D.H. was supported in part by the
U.S. Department of Energy and in part by the Deutsche
Forschungsgemeinschaft under contract Hu 746/1-3.
 We thank the Ohio Supercomputer Center (OSC) for
the use of their facilities under Grant
No.~PHS229 and the H\"ochstleistungsrechenzentrum (HLRZ) J\"ulich for the
use of their T90 and T3E computers.



\pagebreak


\appendix
\section{The high energy limit of the first order term in the NN 
t-matrix}

To evaluate the first order term  of the multiple scattering expansion
for the total cross section in the high energy limit we consider
\begin{equation}
\label{eqA1}
 X_1 \ \equiv \ _2 \langle \Phi \ | \ (t - t^{\dagger}) \
(1 - P_{23}) \ | \Phi \rangle _2 .
\end{equation}
The completeness relation in the three-nucleon space is given by
\begin{eqnarray}
\label{eqA2}
\nonumber
1 &=& \sum_{m_1 m_2 m_3} \ \sum_{\nu_1 \nu_2 \nu_3} \ 
\int d^3p \ d^3q\\
[12pt] 
&&| \ {\bf p} \  m_2 m_3 \ \nu_2 \nu_3 \ > \ | 
{\bf q} \ m_1 \nu_1 \ > \
< {\bf p} \  m_2 m_3 \ \nu_2 \nu_3 \ | \ <{\bf q} \  m_1 \nu_1 \ |
\end{eqnarray}   

Here ${\bf p}$ and ${\bf q}$ are standard Jacobi momenta, $m_i$ the
spin magnetic quantum numbers and $\nu _i$ the corresponding isospin
quantum numbers of the three nucleons. In that notation the channel
state $| \Phi  \rangle_2$ 
reads
\begin{equation}
\label{eqA3} 
\ | \Phi  >_2 \ = \ |  \varphi_d m_d \nu_d \ > \ 
| {\bf q}_0 \ m_N \nu_N  >.
\end{equation}   
We also need a change of basis   

\begin{eqnarray}
\label{eqA4}
\nonumber
&&_2 < {\bf p} \ m_3 m_1 \ \nu_3 \nu_1 \ {\bf q} \ m_2 \nu_2  | 
{\bf p}^{\;'} \ m_2' m_3' \ \nu_2' \nu_3' \ {\bf q}{\;'} \ m_1 ' \nu_1'>_1 \\
&&= \ \delta ({\bf p}^{\;'} - {1 \over 2} \ {\bf q}^{\;'} - {\bf q}) \
\delta ({\bf p} + {\bf q}^{\;'} +  \ {1 \over 2} \ {\bf q})
\ \prod^3_{i=1} \ \delta_{m_i m_i'} \ \prod^3_{i=1} \ \delta\nu_i \nu_i'
\end{eqnarray} 
and the explicit expression for the deuteron state
\begin{eqnarray}
\label{eqA5}
\nonumber
< {\bf p} \ m_3 m_1 \ \nu_3 \nu_1  |  \varphi_d  > &=& \\
\nonumber
& &  {(-)^{{1 \over 2} - \nu_3} \over \sqrt{2}} 
\ \delta_{\nu_{1}, - \nu_{3}} \ \sum_{lm_{l}} \
(l 11, m_l m_d - m_l) \\
 && Y_{lm_{l}} \ ({\hat p}) \ ({1 \over 2} \ {1 \over 2} 1,
	     m_3 \ m_1 \ m_d - m_l) \ \varphi_l  (p)
    \end{eqnarray}

\noindent
As an immediate step one finds after a suitable substitution of 
integration variables
\begin{eqnarray}
    \label{eqA6}
    \nonumber
 _2 <  \Phi  |  t - t^{\dagger} |\Phi >_2 
  &=& \sum_{m_1m_3} \sum_{m_3'} \sum_{\nu_1 \nu_3} \sum_{\nu_3'}  
   \int  d^3 z < \varphi_d  | -{\bf z} \ m_3  m_1  \nu_3 \nu_1  > \\
\nonumber
 & &   < {3 \over 4} {\bf q}_0 + {1 \over 2} {\bf z} 
    m_N m_3  \nu_N \nu_3  | t - t^{\dagger} | 
    {3 \over 4}  {\bf q}_0  +  {1 \over 2}   {\bf z} 
    m_N m_3' \nu_N \nu_3' > \\
\nonumber
& &  < - {\bf z}  m_3'm_1  \nu_3' \nu_1 | \varphi_d >
    \end{eqnarray}     

\noindent
    If the projectile momentum ${3 \over 4} \ {\bf q}_0$ is much larger
than typical momenta ${\bf z}$ occurring in the deuteron state, one can 
take the NN t-matrices out of the integral and obtains

\begin{eqnarray}
\label{eqA7}
    \nonumber
_2< \Phi | t - t^{\dagger} | \Phi >_2 
   & \stackrel{{3 \over 4} q_0 \to \infty}{\longrightarrow}& \\
\nonumber
  & &  \sum_{m_1 m_3} \sum_{m_3'} \sum_{\nu_1 \nu_3} \sum_{\nu_3'} 
   < {3 \over 4} {\bf q}_0  m_N m_3  \nu_N \nu_3  | 
 t-t^{\dagger} | {3 \over 4} {\bf q}_0  m_N m_3'  \nu_N \nu_3'> \times \\
 & & \int d^3 z \ < \varphi_d |- {\bf z}  m_3 m_1  \nu_3 \nu_1  > 
    \ < -{\bf z}  m_3' m_1 \nu_3' \nu_1  |  \varphi_d >.
    \end{eqnarray}  

\noindent
Here the energy argument of the $t$ matrices is given by
 \begin{eqnarray}
 \label{eqA8}
  \nonumber
  \varepsilon \ &\equiv& E - {3 \over 4m}  ( - {1 \over 2} {\bf q}_0 +
    {\bf z})^2 \\
   \nonumber
   &=& \varepsilon _d + {3 \over 4m}  {\bf q}_0^{\;2} 
    - {3 \over 4m} 
    ({1 \over 4} {\bf q}_0^{\;2} - {\bf q}_0  \cdot {\bf z} +
     {\bf z}^{\;2})\\
    \nonumber
    &=& \varepsilon _d + {1 \over m} ({3 \over 4} {\bf q}_0)^2 
    + {3 \over 4m}  {\bf q}_0  \cdot {\bf z} - {3 \over 4m}  
{\bf z}^{\;2}\\
&\stackrel{{3 \over 4} q_0 \to \infty}{\longrightarrow} &
{1 \over m} ({3 \over 4m}  {\bf q}_0)^2 .
\end{eqnarray} 

\noindent
Thus in the high energy  limit we only encounter on-shell NN
t-matrices.

\noindent
Using Eq.(\ref{eqA5}) we find
\begin{eqnarray}
\label{eqA9}
\nonumber  
_2 <\Phi | t - t^{\dagger}  |  \Phi >_2 & \longrightarrow &
\sum_{m_1m_3} \sum_{\nu_1}
< {3 \over 4} {\bf q}_0  m_N m_3  \nu_{N},-\nu_1 | 
t - t^{\dagger} | {3 \over 4} {\bf q}_0  m_N m_3 
\nu_{N}, - \nu_1 > \\
\nonumber
& & {1 \over 2}  \sum_l  \int^\infty_0 \ dp \ p^2  \varphi_l^2 (p) \
C(l11, m_d - m_3 - m_1, m_3 + m_1)^2\\
&& \ \ \ \ \ \ \ \ \ \ \ \ \ \ \ \ \ \ C({1 \over 2}  
	 {1 \over 2}  1, m_3 m_1)^2 .
\end{eqnarray} 

\noindent
The next step is to introduce properly antisymmetrized 
NN states in the NN t-matrix. This comes automatically of course, 
since the form $X_1$ contains the operation $(1-P_{23})$. 
Taking this into account we find 
altogether for the expression Eq.~(\ref{eq:15})
\begin{eqnarray}
 \label{eqA10}
  \nonumber  
<\Phi | P(t - t^{\dagger}) P| \Phi > &\longrightarrow & 
   \sum_{m_1m_3} \sum_{\nu_1} \
     _{na}<{3 \over 4} {\bf q}_0 m_N m_3 \nu_{N}, -\nu_1  |
    t - t^{\dagger}| {3 \over 4} {\bf q}_0  m_N m_3 
    \nu_{N}, - \nu_1 >_{na}\\
    \nonumber
  & & \times \sum_l  \int^\infty_0  dp \ p^2 \varphi_l^2 (p) 
    C(l11, m_d - m_3 - m_1, m_3 + m_1)^2\\
    && \ \ \ \ \ \ \ \ \ \ \ \ \ \ \ \ \ \ C({1 \over 2}  
	     {1 \over 2}  1, m_3 m_1)^2 .
    \end{eqnarray}

\noindent
 Here the antisymmetrized and normalized free NN state is given by
\begin{equation}
\label{eqA11}
    |{\bf p}   m_2  m_3 \nu_2 \ \nu_3 >_{na} \ \equiv \
{1 \over \sqrt{2}}  (1-P_{23})  |  {\bf p} m_2 m_3 \ \nu_2 \nu_3 > .
\end{equation}
Finally we use the relation
\begin{equation}
\label{eqA12}
t - t^{\dagger} = -2 \pi i \ {3 \over 4}  q_0  {m \over 2} 
\sum_{m_2'm_3'} \sum_{\nu_2'\nu_3'} 
\int  d{\hat p}
V  | {3 \over 4}  q_0  {\hat p}  m_2' m_3'  \nu_2' \nu_3' >
^{(+)} \ ^{(+)}<{3 \over 4}q_0 {\hat p} m_2' m_3' \nu_2' \nu_3' |V ,
\end{equation}
where the NN potential $V$ is applied to the two-nucleon scattering 
states $|\ldots \rangle^{(+)}$. 
Furthermore,  the definition of the total NN cross section initiated 
by fixed magnetic spin and isospin quantum numbers is  given by
\begin{equation}
\label{eqA13}
\nonumber 
\sigma_{\nu_N,-\nu_1}^{{\rm tot}} (m_N, m_3) = 
(2 \pi)^4 ({m \over 2})^2
\sum_{m_2'm_3'} \sum_{\nu_2'\nu_3'}  \int d{\hat p} 
| _{na} \langle {3 \over 4} {\bf q}_0  m_N m_3  
\nu_{N}, - \nu_1 | t |
{3 \over 4} q_0 {\hat p}  m_2' m_3'  \nu_2' \nu_3' \rangle |^2 .
\end{equation}
This inserted into Eq.~(\ref{eqA10}) 
    and applying the relation given in Eq.~(\ref{eq:4}) to the total
$nd$ cross section we arrive at
  \begin{eqnarray}
  \label{eqA14}
  \nonumber
 \sigma_{Nd}^{tot}  (m_N, m_d  \nu_N) &=& 
    \sum_{m_1 m_3} \sum_{\nu_1} \sigma_{\nu_N,-\nu_1}^{{\rm tot}} 
    (m_N, m_3)\\
&\times & \sum_l \int_ 0^\infty  dp \ p^2 \varphi_l^2 (p) 
    C(l11, m_d - m_3 - m_1, m_3 + m_1)^2
    C({1 \over 2} {1 \over 2}, m_3, m_1)^2 .
\end{eqnarray}
 This is the contribution of the term Eq.~(\ref{eq:15}) to 
 the total $nd$ cross section 
for initially polarized particles and a nucleon species of type $\nu_N$. 
Averaging over the initial state polarizations one can perform the 
summations over $m_d$ and $m_N$ 
 analytically and using the normalization conditions 
for the s- and d-wave parts of the deuteron wave function one ends up 
for a neutron induced process with
\begin{equation}
 \label{eqA15}
\sigma^{{\rm tot}}_{nd}=\sigma_{np}^{{\rm tot}}+\sigma_{nn}^{{\rm tot}}
.
\end{equation}

\section{The high energy limits of the second order terms in the NN
t-matrix}

To evaluate the second order term of the multiple scattering expansion
for the total $nd$ cross section in the high energy limit we consider
\begin{equation}
\label{eqB1}
 X_2  \equiv  _2 \langle \Phi | (t-t^{\dagger}) \ G_0 \ t_3  | \Phi
\rangle_1
\end{equation} 
Corresponding steps to the ones carried out in Appendix A lead to
\begin{eqnarray}
\label{eqB2}
\nonumber
X_2&=& \sum_{m_1 m_3  \nu_1 \nu_3}  \sum_{m_2' m_3'  \nu_2' \nu_3'} 
\sum_{m_2''  \nu_2``} 
 \int d^3 z \ \int d^3 z{\;'}  < \varphi_d 
 | - {\bf z}  m_3 m_1  \nu_3 \nu_1 > \\
\nonumber
& \times & {< {3 \over 4} {\bf q}_0 + {1 \over 2} {\bf z} \ m_2^0 m_3 
 \nu_2^0\nu_3 | t-t^{\dagger} | {3 \over 4} {\bf q}_0 - {1 \over 2}
 {\bf z} - {\bf z}{\;'} m_2'm_3'  \nu_2' \nu_3' >   
\over - |  \epsilon_d  | - {1 \over m} 
(z^2 + z^{'2} + {\bf z} \cdot {\bf z}{\;'}) + 
{3 \over 2m} ({\bf z} +  {\bf z}{\;'}) \cdot  {\bf q}_0 + i \epsilon}\\
\nonumber
 & \times & <-{3 \over 4} {\bf q}_0 + {\bf z} + {1 \over 2} {\bf z}{\;'}  
m_1 m_2' \ \nu_1 \nu_2'  | \ t \ | {3 \over 4}  {\bf q}_0  
+ {1 \over 2} {\bf z}{\;'} m_1^0 m_2''  \nu_1^0 \nu_2'' >\\
& & < - {\bf z}{\;'} m_2'' m_3'  \nu_2'' \nu_3'  | \varphi_d > .
\end{eqnarray} 
Again we assume that ${3 \over 4} \ q_0$ is much larger than the momenta $z$
and $z'$ in the deuteron state and find
\begin{eqnarray}
\label{eqB2a}
\nonumber
 X_2 & \longrightarrow&  \sum_{m_1 m_3  \nu_1 \nu_3} 
\sum_{m_2' m_3'  \nu_2' \nu_3'}  \sum_{m_2''   \nu_2 ''} 
 < {3 \over 4} {\bf q}_0  m_2^0 m_3  \nu_2^0 \nu_3 |
t-t^{\dagger} | {3 \over 4} {\bf q}_0 \ m_2' m_3' \nu_2' \nu_3' >\\
& \times & <-{3 \over 4} {\bf q}_0 \ m_1 m_2' \nu_1 \nu_2' | \
t \ | {3 \over 4} {\bf q}_0  m_1^0m_2'' \nu_1^0 \nu_2'' > \cdot
{\hat I} ,
\end{eqnarray}
with
\begin{equation}
\label{eqB3}
{\hat I} \ \equiv \ \int d^3 p \int d^3 p{\;'} 
{<\varphi_d |- {\bf p} m_3 m_1  \nu_3 \nu_1 >  
<-{\bf p}{\;'} m_2''m_3' \nu_2'' \nu_3' | \varphi_d > \over
- | \epsilon_d |-{1 \over m} ({\bf p}^{\;2} + {\bf p}^{{\;' 2}}
+ {\bf p} \cdot {\bf p}{\;'}) + {3 \over 2m}
 ({\bf p} + {\bf p}{\;'}) \cdot  {\bf q}_0 + i \epsilon} .
\end{equation}
For the extraction of the asymptotic limit  as
${3 \over 4} |{\bf q}_0 |  \to \infty$ 
we refer to a previous study \cite{ref3} 
carried out for 3 bosons. We use Eq.~(\ref{eqA5}) and 
use the deuteron wave function given in configuration space.
After some algebra one arrives at 
\begin{eqnarray}
\label{eqB4}
\nonumber
{\hat I} &\longrightarrow&  - {i \ m \over 3q_0}  2 \pi^2 (-)^
{{1 \over 2} - \nu_3} (-)  ^{{1 \over 2} - \nu_2'} \ 
\delta_{\nu_1, - \nu_3} \  \delta_{\nu_3', - \nu_2'}\\
\nonumber
& \times & \sum_{l} C(l 11, 0 m_d) ({1 \over 2} {1 \over 2} 1, m_3 m_d-m_3) 
\sqrt{2l+1}\\
\nonumber
& \times &\sum_{l'} C(l'11, 0 m_d) ({1 \over 2} {1 \over 2} 1, m_2' m_d - m_2')
\sqrt{2l'+1}\\
& \times & i^{l'-l} \int_0^\infty dr \ \varphi_l (r)    \varphi_{l'} (r) \
\delta_{m_1, m_d-m_3} \ \delta_{m_3', m_d - m_2'} .
\end{eqnarray} 
The second term in Eq.~(\ref{eq:17}) can be evaluated 
analogously and we end up with the intermediate result
\begin{eqnarray}
\label{eqB5}
\nonumber
\langle \Phi | P (t-t^{\dagger}) P  G_0 tP | \Phi \rangle &=& 
 - {2im \over 3 q_0} \ (2 \pi)^2 \ \sum_{m_3 \nu_3} \
 \sum_{m_2' \nu_2'} \ \sum_{m_2'' \nu_2''}\\
\nonumber
&&< {3 \over 4} {\bf q_0}  m_2^0 m_3  \nu_2^0 \nu_3  | t-t^{\dagger} | 
{3 \over 4} {\bf q_0}  m_2' m_d -m_2'', \nu_2', - \nu_2'' >_{na}\\
\nonumber
&&< - {3 \over 4} {\bf q_0}  m_d - m_3, m_2',-\nu_3, \nu_2'  | t | 
{3 \over 4} {\bf q_0} m_1^0m_2''  \nu_1^0 \nu_2'' >_{na} 
(-)^{{1 \over 2} - \nu_3}  (-)^{{1 \over 2} - \nu_2''}\\     
\nonumber
&& \sum_{l} \ \sqrt{2l + 1} \ C(l11, 0 m_d) \ ({1 \over 2} {1 \over 2} 1,
m_3 m_d-m_3)\\
\nonumber
&& \sum_{l'} \ \sqrt{2l' + 1} \ C(l'11, 0 m_d) \ ({1 \over 2} {1 \over 2} 1,
m_2'' m_d-m_2'')\\
&&i^{l'-l} \ \int^\infty_0 \ dr \ \varphi_{l'}(r) \ \varphi_{l}(r) .
\end{eqnarray} 
We encounter only forward or backward NN scattering amplitudes, which 
induces connections between the magnetic spin quantum numbers occurring 
in the t-matrices (put ${\hat q}_0 = {\bf z}$). Further we sum over 
the initial spin magnetic quantum numbers $m_d$ and $m_N$.  
As an example of these summations we show the result 
\begin{eqnarray}
\label{eqB6}
\nonumber
& &\sum_{m_N m_d} \{ \langle \Phi | P(t-t^{\dagger}) P \ G_0 \ tP|\Phi
\rangle 
- \ complex \ conjugate  \} = \\
\nonumber
& & i{3q_0 \over 2m} {1 \over 2 \pi^4} \sum_{l} \sqrt{2l + 1} 
 \sum_{l'}  \sqrt{2l' + 1}  i^{l'-l}  \int^\infty_0 dr \ \varphi_l(r)  
\varphi_{l'}(r) \\
\nonumber
&&  \bigg[ 2C(l11,01) C(l'11,01)
 \{ \sigma^{tot}_{NN} ({1 \over 2} {1 \over 2} np) + 
\sigma^{tot}_{NN} ({1 \over 2} {1 \over 2} nn) + 
\sigma^{tot}_{NN} ({1 \over 2} - {1 \over 2} np) 
\sigma^{tot}_{NN} ({1 \over 2} - {1 \over 2} nn) \} \\ 
\nonumber
&& +C(l11,00)C(l'11,00)
 \{ \sigma^{tot}_{NN} ({1 \over 2} {1 \over 2} np) 
 \sigma^{tot}_{NN}  ({1 \over 2}- {1 \over 2} nn) + 
\sigma^{tot}_{NN}  ({1 \over 2}  - {1 \over 2} np) 
\sigma^{tot}_{NN}  ({1 \over 2} {1 \over 2} nn) \} \bigg] \\
\nonumber 
&&-i {8m \over 3q_0} \ (2 \pi)^2  \sum_l  \sqrt{2l+1} \ 
\sum_l  \sqrt{2l'+1} \ i^{l'-l} \
\int^\infty_0 dr \ \varphi_l(r)   \varphi_{l'}(r)\\
\nonumber
&&  \bigg[ \  2C(l11,01)C(l'11,01) \\
\nonumber
&& \{ (Im < {3\over 4}{\bf q_0} 
{1 \over 2} {1 \over 2} np | t | 
  {3\over 4}{\bf q_0} {1 \over 2}{1 \over 2} pn >_{na})^2
+ (Im <  {3\over 4}{\bf q_0} 
{1 \over 2} - {1 \over 2} np \ | \ t \ | \
  {3\over 4}{\bf q_0} {1 \over 2}  - {1 \over 2} pn >_{na})^2 \}\\
\nonumber
&& + C(l11,00) C(l'11,00)\\
\nonumber
&&\{2 (Im <  {3\over 4}{\bf q_0} 
{1 \over 2}   {1 \over 2} np  | t |
 {3\over 4}{\bf q_0}  {1 \over 2}   {1 \over 2} pn >_{na}) (Im 
<  {3\over 4}{\bf q_0} 
{1 \over 2}  - {1 \over 2} pn  | t |
  {3\over 4}{\bf q_0} {1 \over 2}  - {1 \over 2} np >_{na})\\
&& + (Im <   {3\over 4}{\bf q_0} 
{1 \over 2}  - {1 \over 2} np  | t | -
  {3\over 4}{\bf q_0} {1 \over 2} {1 \over 2} pn >_{na})^2\\
\nonumber
&& -2 (Im < {3\over 4}{\bf q_0}
 {1 \over 2}  - {1 \over 2} np | t| -
 {3\over 4}{\bf q_0} {1 \over 2} {1 \over 2} np >_{na})
 (Im < {3\over 4}{\bf q_0} 
{1 \over 2}  - {1 \over 2} nn  |t| -
  {3\over 4}{\bf q_0} {1 \over 2} {1 \over 2} nn >_{na})\}.
\end{eqnarray}
The last term in Eq.~(\ref{eq:7}) can be treated analogously to 
Eq.(\ref{eqB1}) resulting in
\begin{eqnarray}
\label{eqB7}
\nonumber
 \langle \Phi | P t^{\dagger} P  (G_0 - G_0^*) tP | \Phi \rangle
&\to& - 8 \pi i  \sum_{m_1 m_3 \nu_1 \nu_3}  
\sum_{m_{2'}m_{3'} \nu_{2'} \nu_{3'}}  
\sum_{m_{2''} \nu_{2''}} \\
\nonumber 
&& < {3 \over 4} {\bf q_0}  m_2^0 m_3  \nu_2^0 \nu_3 | t^{\dagger} |
{3 \over 4} {\bf q_0}  m_2' m_3', \nu_2' \nu_3' >_{na}\\
& \times &<-{3 \over 4} {\bf q_0}  m_1 m_2'  \nu_1 \nu_2' | t |
{3 \over 4} {\bf q_0}  m_1^0 m_2'' \nu_1^0 \nu_2'' >_{na} \  {\tilde I},
\end{eqnarray}
where
\begin{eqnarray}
\label{eqB8}
\nonumber
{\tilde I}& =& \int d^3 p \int d^3 p{\;'}  
< \varphi_d | -{\bf p} m_3 m_1  \nu_3 \nu_1 >
< -{\bf p}^{\;'} m_2' m_3'  \nu_2' \nu_3'| \varphi_d >\\
\nonumber
&&\delta(- | \epsilon _d | - {1 \over m} ( {\bf p}^{\;2} +  
{\bf p}^{\; '2} + {\bf p}{\;'} \cdot {\bf p}) + 
{3 \over 2m} {\bf q}_0 \cdot ({\bf p} + {\bf p}{\;'})).
\end{eqnarray}
It is easily shown that in the high energy limit
\begin{equation}
{\tilde I} = - {1 \over i \pi} \ {\hat I}.
\end{equation}
It remains to perform the summations over the spin- and isospin quantum 
numbers. Then using the connection  to 
the total $nd$ cross section, Eq.~(\ref{eq:4}), 
the lengthy expression Eq.~(\ref{eq:19}) results.


\newpage

\begin{table}

\begin{tabular} {|c|c|c|c|c|c|}
 $E_{lab}$ (MeV)&\multicolumn{5}{c|}{ ${\sigma}_{nd}^{{\rm tot}}$ (mb) }\\
\hline
 &$j_{max}=1$&$j_{max}=2$&$j_{max}=3$&$j_{max}=4$&$j_{max}=5$\\
\hline
    10.0&954.9&1038.0&1035.2&1037.3&1036.7 \\
\hline
    60.0&141.0&177.8&177.0&177.4&177.3\\
\hline
    140.0&43.0&71.6&73.2&74.7&74.6\\
\hline
    200.0&31.7&55.8&58.3&60.5&60.4\\
\hline
    260.0&29.7&49.4&52.2&54.7&54.7\\
\hline
    300.0&29.9&46.9&49.7&52.3&52.4\\
\end{tabular}
\caption{\label{tab1} The convergence of the calculated 
total $nd$ cross section 
with increasing $j_{max}$ at selected energies. As NN interaction 
 the CD-Bonn potential has been used. }

\vspace{10mm}

\begin{tabular} {|c|c|c|c|c|}
 $E_{lab}$ (MeV)&\multicolumn{4}{c|}{ ${\sigma}_{nd}^{tot}$ (mb)  }\\
\hline
 &AV18&CD-Bonn&NijmI&NijmII\\
\hline
    10.0&1039.0&1035.2&1038.0&1038.3 \\
\hline
    100.0&97.7&99.3&98.2&97.5\\
\hline
    140.0&72.2&73.2&72.5&71.8\\
\hline
    200.0&57.8&58.3&57.9&57.4\\
\hline
    300.0&49.5&49.7&49.6&49.3\\
\end{tabular}
\caption{\label{tab2} Stability of the calculated total $nd$ cross section 
 ${\sigma}_{nd}^{{\rm tot}}$ under the exchange of the NN potentials. 
The calculations were performed with $j_{max}=3$. }

\newpage

\begin{tabular} {|c||c|c|c|c||c|c|c|c|}
 $E_{lab}$ (MeV)&\multicolumn{8}{c|}{ ${\sigma}_{nd}^{tot}$ (mb)  }\\
\hline
  &\multicolumn{4}{c|}{Faddeev Calculation}&\multicolumn{4}{c|}
{High Energy Expansion} \\
\hline
 &full FC& 1.ord.& 2.ord.& 1.+2.ord. & 
   $\sigma_{np}+ \sigma_{nn}$& $\sigma_{I}^{(2)}$ & $\sigma_{II}^{(2)}$ & sum \\
\hline
 100. &100.95&108.27& 1.91&110.18&105.60& 20.62& -6.54&119.67\\
 140. & 74.85& 78.62& 0.08& 78.69& 80.23& 12.20& -3.96& 88.45\\
 200. & 60.55& 62.74&-0.86& 61.89& 65.86&  6.28& -2.80& 69.34\\
 300. & 52.30& 54.05&-1.52& 52.53& 58.14&  2.38& -2.28& 58.24\\
\end{tabular}
\caption{\label{tab3} Comparison of Faddeev results with the  asymptotic
expansion ones. The calculations were performed throughout  with the
CD-Bonn potential and $j_{max}=4$. The terms  $\sigma_{I}^{(2)}$ and
$\sigma_{II}^{(2)}$ are from
Eq. (2.19) . For more explanation of the different
terms see Sec. II and III. }

\end{table}

\pagebreak

\noindent

\noindent

\noindent

\noindent
 
\noindent




\end{document}